# Solid-state laser cooling of Yb$^{3+}$-doped KY$_3$F$_{10}$ to 145 K

Luca Koldeweyh,[†] Stefan Püschel,[†,‡] Zoe Liestmann, Hiroki Tanaka[*]

*Leibniz-Institut für Kristallzüchtung (IKZ), Max-Born-Str. 2, 12489 Berlin, Germany*
[†]*These authors contributed equally to this work.*
[‡]*Current affiliation: Fraunhofer Institute of Optronics, System Technologies and Image Exploitation (IOSB), Gutleuthausstr. 1, 76275 Ettlingen, Germany*
[*]*hiroki.tanaka@ikz-berlin.de*

**We report laser cooling of Yb$^{3+}$-doped KY$_3$F$_{10}$ (Yb:KY$_3$F$_{10}$) driven by a 100-W, 1020-nm pump source. Despite pumping at a non-optimal wavelength, high-quality KY$_3$F$_{10}$ crystals doped with 3% and 7% Yb were cooled to 145 K and 151 K, respectively, in a double-pass pump configuration. These results establish Yb:KY$_3$F$_{10}$ as an attractive laser-cooling medium competitive with Yb:YLF, the state-of-the-art laser-cooling material for optical cryocoolers. The observed cooling performance and spectroscopic characteristics suggest that lower cryogenic temperatures may be achieved through pump-wavelength optimization, enhanced pump absorption, and reduced radiative heating.**

Solid-state laser cooling [1–4], also known as optical refrigeration, provides a cooling solution capable of reaching temperatures below the limit of thermoelectric coolers of ~170 K, without using cryogenic fluids. This vibration-free, all-solid-state cooling technology is attractive for applications in space [5] and high-precision metrology [6]. The absence of cryogenic fluids is particularly beneficial for space-based systems, where leakage risks and fluid management are critical concerns.

Solid-state laser cooling is achieved through the anti-Stokes fluorescence process, where the mean fluorescence photon energy exceeds the excitation photon energy. This process removes thermal energy by annihilating phonons and releasing it through the emission of higher-energy fluorescence photons, resulting in cooling the solid. Efficient cooling requires optically active media with high fluorescence quantum efficiencies and low impurity-induced background absorption; otherwise, non-radiative relaxation processes may suppress the cooling effect.

Owing to these requirements, trivalent ytterbium (Yb$^{3+}$)-doped crystals have been extensively studied as laser-cooling media. Yb$^{3+}$ exhibits fluorescence quantum efficiencies near unity in many oxide and fluoride hosts [7]. Although laser cooling has been observed in various materials [2,8], high-power cooling experiments targeting the lowest achievable temperatures have been largely limited to Yb$^{3+}$:LiYF$_4$ (Yb:YLF) [9–12] and its isomorph LiLuF$_4$ (Yb:LLF) [13].

Low crystal-field hosts, including fluorides, are advantageous because their smaller Stark level splitting helps to maintain cooling efficiency at low temperatures [14]. In addition, fluoride crystals typically possess lower refractive indices (≈1.4–1.5) [15] than common oxide hosts, *e.g.*, ≈1.82 for Y$_3$Al$_5$O$_{12}$ (YAG) at a wavelength of 1 µm [16], improving the fluorescence escape efficiency by reducing total internal reflection. To date, the lowest temperature achieved by laser cooling is 87 K using Yb:YLF [12], and optical cryocooler prototypes were developed based on this material, with payload temperatures below 125 K [19].

Despite the success with Yb:YLF, its widespread use has been partly driven by the availability of high-quality crystals rather than the intrinsic suitability over other materials. In our previous work, we identified Yb:KY$_3$F$_{10}$ as a more suitable candidate for optical cryocoolers, based on its spectroscopic properties [19].

In this work, we demonstrate laser cooling in Yb:KY$_3$F$_{10}$ using a high-power pump source. Despite the pump wavelength not being optimized to reach the lowest achievable temperature, Yb(3%):KY$_3$F$_{10}$ was cooled down to 145 K in a double-pass configuration without suppression of radiative heating from the surroundings.

We grew two KY$_3$F$_{10}$ single crystals (cubic structure, $F\bar{m}3m$ space group, No. 225 ITA) doped with 3.0% and 7.3% of Yb ions in the melt, corresponding to the chemical formulae K(Yb$_{0.03}$Y$_{0.97}$)$_3$F$_{10}$ and K(Yb$_{0.073}$Y$_{0.927}$)$_3$F$_{10}$, respectively. The starting materials, KF powder (6N for 3.0% and 4N for 7.3% crystals, Fox-Chemicals) and YF$_3$ crystalline granules (5N5, AC Materials) were mixed in a glovebox with argon atmosphere and subsequently purified by hydrofluorination, wherein the mixed compounds were heated under a flow of argon mixed with HF gas. The fluorinated materials and fluorinated YbF$_3$ crystalline granules (5N, AC materials) were weighed into a glassy carbon crucible (HTW Hochtemperatur Werkstoffe). The two crystals were grown in an induction-heated Czochralski puller in 6N CF$_4$ atmosphere, along the crystallographic [100]-axis at a growth rate of 1 mm/h and a rotation speed of 10 rpm.

From each crystal boule, we prepared a rectangular-shaped sample, which was six-facet-polished for the laser-cooling experiments. The actual Yb$^{3+}$ doping concentrations in the samples were determined to be 3.0% and 7.0% from their absorption coefficients using the previously determined absorption cross sections [20]. Hereafter, these samples are denoted as Yb(3%):KY$_3$F$_{10}$ and Yb(7%):KY$_3$F$_{10}$. Their dimensions are 2.8×3.2×9.7 mm$^3$ and 2.7×3.0×10.7 mm$^3$, respectively.

We characterized these samples using laser-induced thermal modulation spectroscopy (LITMoS) [21]. The details of our experimental setup are described in Ref. [21]. **Figure 1** presents the LITMoS results, showing the laser-induced temperature change normalized by the absorbed power as a function of the pump wavelength, $\Delta T/P_{abs}(\lambda)$. The low available pump power around 1065 nm caused the high uncertainties in this range. The LITMoS results suggested external quantum efficiencies (EQE) $\eta_{ext}$ of 99.5±0.1% and 99.0±0.1%, and background absorption coefficients $\alpha_b$ of $(7\pm1)\cdot10^{-5}$ cm$^{-1}$ and $(3\pm1)\cdot10^{-5}$ cm$^{-1}$, for Yb(3%):KY$_3$F$_{10}$ and Yb(7%):KY$_3$F$_{10}$, respectively. We determined the EQE and the background absorption coefficients based on two zero-crossing wavelengths, $\lambda_{X1}$ and $\lambda_{X2}$, where neither heating nor cooling was observed [21]. In this analysis, we used room-temperature mean fluorescence wavelengths $\lambda_f$ of 993.0 nm and 997.5 nm, calculated by Monte Carlo fluorescence ray tracing simulations [22], considering their dimensions and Yb$^{3+}$ concentrations. The Yb(3%):KY$_3$F$_{10}$ sample exhibited an EQE higher by ≈0.5%pt. compared with our previous work [19,22]. The lower EQE in Yb(7%):KY$_3$F$_{10}$ may be due to a stronger impact of foreign rare-earth impurities, primarily Tm$^{3+}$ and Ho$^{3+}$ [23], associated with the higher Yb$^{3+}$ concentration. The high Yb$^{3+}$ concentration enhances energy migration among Yb$^{3+}$ ions, thereby promoting the energy transfer to these impurities. The background absorption coefficients below 10$^{-4}$ cm$^{-1}$ are among the lowest values ever reported for Yb$^{3+}$-doped crystals, as the typical values for Yb:YLF are on the order of 10$^{-4}$ cm$^{-1}$ at room temperature [9,10,19,24]. This indicates that, compared with YLF, the structure of KY$_3$F$_{10}$ tends to incorporate fewer impurities that cause background absorption. Despite higher-purity KF (6N) being used for Yb(3%):KY$_3$F$_{10}$, whereas 4N KF was used for Yb(7%):KY$_3$F$_{10}$, Yb(3%):KY$_3$F$_{10}$ exhibited higher background absorption. This suggests that KF is not the dominant source of impurities causing background absorption.

The determined EQEs and background absorption coefficients allow us to calculate the cooling efficiency $\eta_c(\lambda)$, which is proportional to the measured quantity $\Delta T/P_{abs}(\lambda)$ [25]:

$$\eta_c(\lambda, T) = \eta_{ext}\left[\frac{\alpha_r(\lambda,T)}{\alpha_r(\lambda,T)+\alpha_b}\right]\frac{\lambda}{\lambda_f(T)} - 1, \quad (1)$$

where $\alpha_r$ is the resonant absorption due to Yb$^{3+}$. The black solid lines in **Fig. 1** are calculated using Eq. (1), suggesting maximum room-temperature cooling efficiencies of 4.0% and 4.5% at pump wavelengths of ≈1047 nm and ≈1065 nm for Yb(3%):KY$_3$F$_{10}$ and Yb(7%):KY$_3$F$_{10}$, respectively.

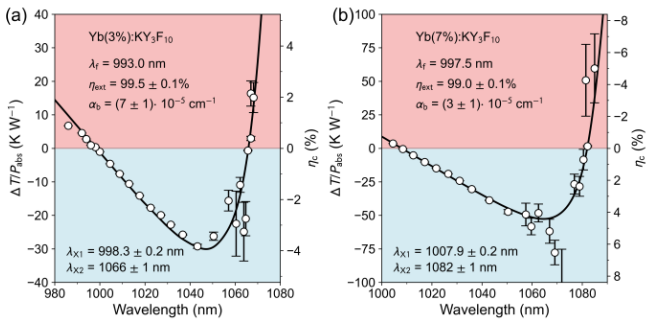

Fig. 1. Laser-induced thermal modulation spectroscopy (LITMoS) results for Yb(3%):KY$_3$F$_{10}$ (a) and Yb(7%):KY$_3$F$_{10}$ (b). The symbols are experimentally measured values, and the black solid curves are calculated using Eq. (1) using $\lambda_f$, $\eta_{ext}$, and $\alpha_b$ denoted in the figures. The observed zero-crossing wavelengths $\lambda_{X1}$ and $\lambda_{X2}$ are also denoted in the figures.

We subsequently performed high-power laser cooling experiments. **Figure 2** shows the setup in the single- and double-pass configurations using a 1020-nm Yb fiber laser (IPG Photonics, YLR-100-1020-LP-WC) delivering up to 100 W. The laser-cooling samples were placed inside a stainless-steel vacuum chamber kept at 295 K by water cooling. The sample was supported by two bare glass fibers (SM980, 125 µm diameter) suspended over a copper holder to minimize conductive heat exchange with the surroundings. The output of the Yb fiber laser was shaped by a telescope and guided to the sample through an optical isolator (Newport, ISO-FRDY-05-1030-N). The collimated beam diameter in the sample was measured to be ~1 mm. Pumping with such a large collimated beam was to minimize absorption bleaching, thus maximizing the pump absorption. In the single-pass configuration (**Fig. 2a**), we measured the transmitted power using a thermal power meter (Ophir Optronics Solutions, 150A) during the cooling experiments. In the double-pass configuration (**Fig. 2b**), a plane high-reflectivity (HR) mirror reflected the transmitted beam back into the sample. To prevent additional absorption bleaching by the back-propagating beam, the overlap between the counter-propagating beams in the sample was minimized. During the measurements, the vacuum chamber was evacuated using a turbo-molecular pump to a pressure below 10$^{-6}$ mbar. We applied differential luminescence thermometry (DLT) [19] to determine the temperature of the cooled samples. For this purpose, the fluorescence was collected via a multimode fiber (600 µm core diameter), guided through a fiber feedthrough, and recorded with a fiber-coupled spectrometer (Ocean Optics, HR4000).

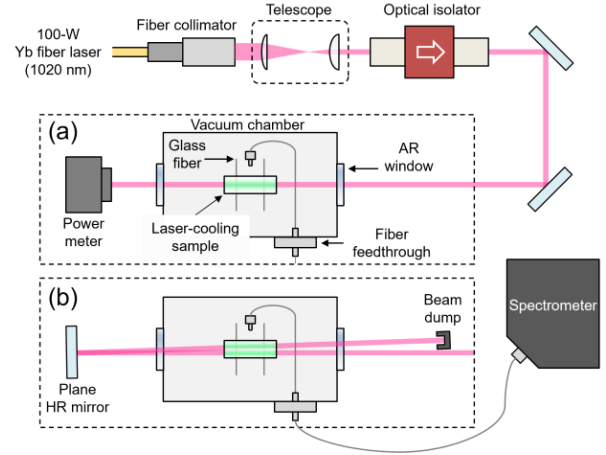

Fig. 2. Laser-cooling setup in single- (a) and double-pass (b) pumping configurations with a Yb fiber laser emitting at 1020 nm.

For temperature calibration in DLT, we recorded the fluorescence spectra $S(\lambda, T)$ of both samples from room temperature to 50 K in 10 K increments, using a closed-cycle helium cryostat (Applied Research Systems, DE204). In these measurements, the samples were pumped by a continuous-wave Ti:sapphire laser (SolsTiS, M Squared Lasers) tuned to 900 nm. To minimize the laser-induced local heating, the pump power was limited to ≈60 mW. These calibration measurements were performed under pump and detection settings similar to those in the laser-cooling setup for minimizing the uncertainty in the determined temperature. The DLT calibration datasets were prepared separately for single- and double-pass configurations. For the double-pass configuration, to closely mimic the pump geometry in the laser-cooling setup, the spectra recorded at two different pump beam positions were averaged and used. The DLT calibration data $S_{\text{DLT}}(T, T_0)$ were calculated according to [19]:

$$S_{\text{DLT}}(T, T_0) = \int_{\lambda_1}^{\lambda_2} \left| \frac{S(\lambda,T)}{\int S(\lambda,T)d\lambda'} - \frac{S(\lambda,T_0)}{\int S(\lambda,T_0)d\lambda'} \right| d\lambda. \quad (2)$$

Here, $T_0$ is the starting temperature, set to 295 K. The integral range was chosen as $\lambda_1$=985 nm to $\lambda_2$=1010 nm to minimize reabsorption due to the overlap of absorption and emission spectra, which particularly affects the detected fluorescence spectra below this range. We investigated the impact of pump beam positions on the temperature determined by DLT, and estimated the uncertainty to be ±2 K for the temperature range below 200 K. The use of the narrow integral range was found to reduce this uncertainty.

The high-power laser-cooling results are shown in **Fig. 3**. Besides the two Yb:KY$_3$F$_{10}$ samples, we also characterized a Yb(5%):YLF sample with dimensions of 2.7×3.8×9.7 mm$^3$, $\eta_{\text{ext}}$=99.7±0.1%, and $\alpha_b$=(7±1)·10$^{-5}$ cm$^{-1}$ in the same setup. Using the maximum power of the Yb fiber laser, corresponding to an incident pump power of 89 W, the Yb(3%):KY$_3$F$_{10}$ and Yb(7%):KY$_3$F$_{10}$ samples were cooled from 295 K to 161 K and 160 K, respectively, within ~15 minutes in the single-pass configuration. In the double-pass configuration, they were cooled to 151 K and 145 K, respectively. We attribute the lower temperature achieved with Yb(3%):KY$_3$F$_{10}$ to its higher EQE. At the temperatures of ~150 K, the resonant absorption coefficients are on the order of 10$^{-2}$ cm$^{-1}$, indicating that background absorption has not yet become the limiting factor.

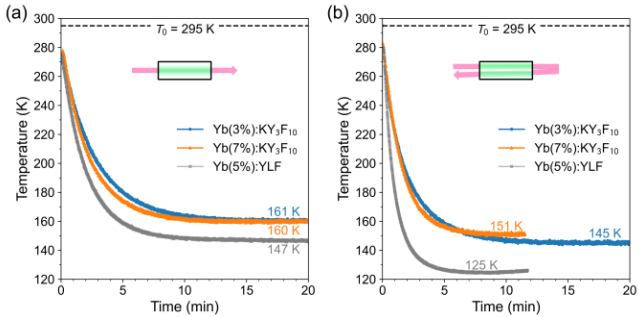

Fig. 3. High-power laser-cooling results of Yb(3%):KY$_3$F$_{10}$ and Yb(7%):KY$_3$F$_{10}$ samples at an incident pump power of 89 W from a Yb fiber laser in single-pass (a) and double-pass (b) pump configuration. For comparison, the results of a Yb(5%):YLF sample are shown. The lowest observed temperatures are denoted in the figures.

Note that the pump wavelength of 1020 nm is not optimized for achieving the lowest temperature in Yb:KY$_3$F$_{10}$, whereas it is the optimum for Yb:YLF. This resulted in faster cooling of the Yb:YLF sample down to 125 K in the double-pass configuration (see **Fig. 3b**). In our previous work, the optimum pump wavelength for cryogenic cooling of Yb(3%):KY$_3$F$_{10}$ was estimated to be below 1013 nm, where the absorption coefficient is an order of magnitude higher than at 1020 nm [25].

**Figure 4a** shows the absorption cross-section spectra of Yb(3%):KY$_3$F$_{10}$, calculated from the emission cross-section spectra [23] using the reciprocity relation (McCumber theory) [26], for temperatures between 60 and 300 K. **Figure 4b** shows the temperature-dependent absorption coefficient at 1013 nm and 1020 nm for Yb(3%):KY$_3$F$_{10}$. At room temperature, the absorption coefficient at 1020 nm is as low as 0.1 cm$^{-1}$ and 0.23 cm$^{-1}$ for 3% and 7% Yb doping, respectively. This resulted in a single-pass absorbed power of ≈8.5 W and ≈18.3 W in the ~10-mm-long samples, whereas the absorption coefficient for Yb(5%):YLF is ≈1 cm$^{-1}$ at this wavelength. At 145 K, the lowest temperature achieved in the Yb(3%):KY$_3$F$_{10}$, the absorption coefficient at 1020 nm decreases to ≈0.01 cm$^{-1}$ (see **Fig. 4b**), leading to an absorbed power of ≈1.8 W and a cooling power of only ≈15 mW in the double-pass configuration. In thermal equilibrium at 145 K (see **Fig. 3b**), the radiative heating power must equal the cooling power. This low heating power is likely due to the higher infrared transmission of KY$_3$F$_{10}$ compared with YLF, which reduces absorption of thermal radiation from the surroundings.

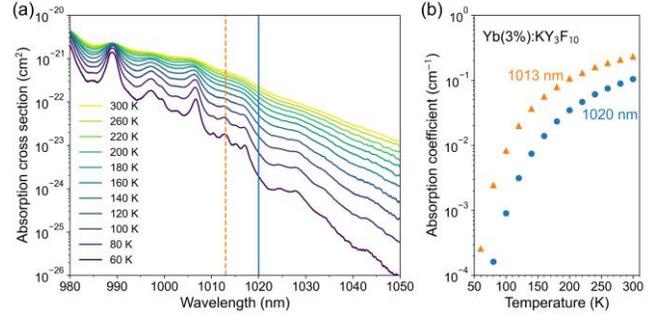

Fig. 4. (a) Absorption cross-section spectra of Yb:KY$_3$F$_{10}$ for temperatures between 60 K and 300 K, calculated from the emission cross-section spectra using the reciprocity relations. The blue solid and orange dashed vertical lines indicate wavelengths of 1020 nm and 1013 nm, respectively. (b) Temperature-dependent absorption coefficient of Yb(3%):KY$_3$F$_{10}$ at 1020 nm and 1013 nm.

**Figure 5** shows the cooling efficiency of the two laser-cooling samples as a function of pump wavelength and temperature, calculated using Eq. (1). The global minimum achievable temperature (MAT) is found to be 105 K at 1013 nm and 102 K at 1017 nm for the Yb(3%):KY$_3$F$_{10}$ and Yb(7%):KY$_3$F$_{10}$ samples, respectively. On the other hand, the MATs at a pump wavelength of 1020 nm are limited to 129 K and 114 K, respectively. Despite the small absorbed power in the Yb(3%):KY$_3$F$_{10}$, the lowest achieved temperature of 145 K is only 16 K above the MAT at 1020 nm. This gap can be further reduced by increasing pump absorption, so cooling power, and reducing the radiative heating from the surroundings. **Figure 5** suggests that a pump wavelength of 1017 nm is promising for cryogenic cooling of both doping concentrations. The slight shift of the pump wavelength from 1020 nm to 1017 nm decreases the MAT by ~20 K.

Note that we considered the background absorption coefficient to remain constant upon cooling, although it has been reported to decrease in Yb:YLF by an order of magnitude when cooled from room temperature to ~100 K [24]. A similar temperature dependency in background absorption can be expected in Yb:KY$_3$F$_{10}$, implying that temperatures below the calculated MATs shown in **Fig. 5** may be achievable. **Figure 5a** also indicates that the MAT of the Yb(3%):KY$_3$F$_{10}$ at a pump wavelength of 1007 nm may decrease to ~70 K by further improving the EQE, as previously discussed [19].

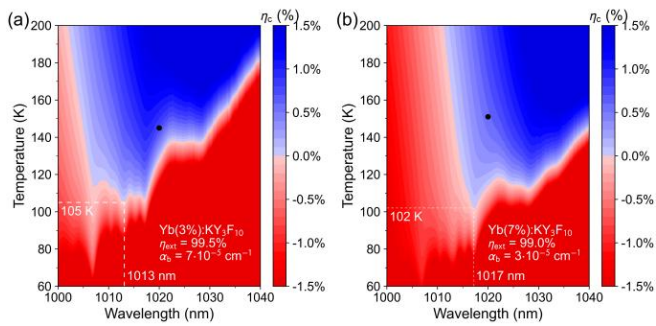

Fig. 5. Cooling efficiency as a function of wavelength and temperature for the Yb(3%):KY$_3$F$_{10}$ (a) and Yb(7%):KY$_3$F$_{10}$ (b) samples, calculated using Eq. (1) with $\eta_{ext}$ and $\alpha_b$ denoted in the figures. These two parameters are considered to be independent of temperature. The black dots indicate the lowest temperatures experimentally achieved in this work at a pump wavelength of 1020 nm.

In conclusion, we experimentally demonstrated that Yb:KY$_3$F$_{10}$ is a promising laser-cooling material for optical cryocoolers. In a simple double-pass pump configuration, KY$_3$F$_{10}$ crystals doped with 3% and 7% Yb were cooled to 145 K and 151 K, respectively, despite a non-optimized pump wavelength. Beyond Yb:YLF and its isomorph Yb:LLF, Yb:KY$_3$F$_{10}$ emerges as an additional material capable of cooling below 150 K. Our results suggest that Yb:KY$_3$F$_{10}$ can reach the cryogenic temperature, defined to be 123 K by the National Institute of Standards and Technology, by optimizing the pump wavelength, enhancing the pump absorption, *e.g.*, using an astigmatic Herriott cell, and by reducing radiative heating using low-emissivity surroundings. Achieving 77 K, the liquid-nitrogen temperature, widely regarded as a key milestone of solid-state laser cooling, could be feasible with further optimization.

**Funding.** Deutsche Forschungsgemeinschaft (520253663); Leibniz-Gemeinschaft (J183/2023).

**Acknowledgment.** The authors acknowledge the technical support by Celine Kapella and John Stahl (both IKZ) in the growth of the fluoride crystals used in this work.

**Disclosures**. The authors declare no conflicts of interest.

**Data availability.** Data underlying the results presented in this paper are available from the authors upon request.